# Bridging spatial and temporal scales of developmental gene regulation

Andrés H. Cardona[1§], Márcia Mesquita Peixoto[1§], Tohn Borjigin[2], Thomas Gregor[1,2,3*]

[1]Department of Stem Cell and Developmental Biology, CNRS UMR3738 Paris Cité, Institut Pasteur, 25 Rue du Docteur Roux, 75015 Paris, France
[2]Lewis-Sigler Institute for Integrative Genomics, Princeton University, Princeton, NJ 08544, USA
[3]Joseph Henry Laboratories of Physics, Princeton University, Princeton, NJ 08544, USA

[§]These authors contributed equally.
[*]Correspondence: thomas.gregor@pasteur.fr

The development of multicellular organisms relies on the precise coordination of molecular events across multiple spatial and temporal scales. Understanding how information flows from molecular interactions to cellular processes and tissue organization during development is crucial for explaining the remarkable reproducibility of complex organisms. This review explores how chromatin-encoded information is transduced from localized transcriptional events to global gene expression patterns, highlighting the challenge of bridging these scales. We discuss recent experimental findings and theoretical frameworks, emphasizing polymer physics as a tool for describing the relationship between chromatin structure and dynamics across scales. By integrating these perspectives, we aim to clarify how gene regulation is coordinated across levels of biological organization and suggest strategies for future experimental approaches.

**Introduction**

The development of complex organisms, from a single cell to multicellular tissues, unfolds over extended periods—days, weeks, or even months. Though this process progresses gradually, it is driven by countless molecular events occurring on much faster and smaller spatiotemporal scales. How information from these rapid, localized events propagates to guide organismal development remains poorly understood. Bridging this gap requires dissecting the intermediate steps and understanding the connections between them.

Transcriptional regulation offers a central framework for exploring how molecular dynamics influence cellular behaviors and tissue-level processes. Chromatin interactions with regulatory effectors, occurring at the nanometer scale within seconds to minutes (Figure 1A), influence and are influenced by larger-scale chromatin reorganization, which spans micrometer distances across minutes to hours (Figure 1B). These layers of regulation collectively guide developmental processes that unfold over longer timeframes, at the tissue and organism scale (Figure 1C). Clarifying how rapid molecular dynamics drive broader developmental changes is essential for resolving the "black boxes" in gene regulation.

At the molecular level, transcription often relies on spatial interactions between gene promoters and cis-regulatory elements (CREs), such as enhancers, located thousands (kb) to millions (Mb) of base pairs apart (Figure 1A). However, the precise spatiotemporal nature of enhancer-promoter (E-P) interactions remains unclear, highlighting the complexity of the three-dimensional (3D) chromatin architecture involved in gene regulation [1]. Advances in high-throughput chromosome conformation capture (3C) technologies have revealed multiple levels of chromatin organization, including loops, topologically associating domains (TADs), compartments (A/B), and chromosome territories (Figure 1A, B) [2]. These chromatin structures are increasingly recognized as key regulators of gene expression during development. For example, in vertebrates, TADs, CCCTC-binding factor (CTCF) insulators, and cohesin-mediated loop extrusion orchestrate the temporal activation of Hox genes, which are critical for body axis formation [3]. Understanding how chromatin structure regulates transcription across scales is crucial for linking localized molecular mechanisms to broader developmental outcomes (Figure 1C).

This review explores how gene regulation operates at three biological scales: (1) the locus scale, connecting transcription effectors, E-P interactions, and transcriptional dynamics; (2) the chromatin scale, emphasizing the random search process involved in long-distance regulation, chromatin structure, and polymeric behavior; and (3) the organism scale, where transcriptional programs and genome organization are coordinated among cell populations and tissues. Bridging these scales remains a significant challenge, requiring complex model systems, high-resolution techniques, and methods capable of capturing "4D" genome-scale data [4]. Current experimental approaches often face trade-offs between spatial or temporal resolution or focus on specific genomic loci, which can obscure the full picture of gene regulation. While this review does not delve into the limitations of these methods, recent publications have discussed their technical biases [1,2,5].

Polymer models offer a promising approach for linking chromatin structure with dynamic regulatory processes. However, discrepancies between experimental observations and theoretical models underscore the need for a more unified framework that integrates transcriptional dynamics and chromatin behavior across scales. In addition, species-specific characteristics must be carefully considered when extrapolating chromatin behavior between model organisms. In this review, we also discuss how polymer models can help bridge spatial and temporal dimensions of gene regulation and the challenges of modelling chromatin, emphasizing the importance of incorporating species-specific factors for a comprehensive understanding of how molecular information is integrated during development.

**Integrating Molecular Dynamics at the Gene Locus**

At the gene locus scale, transcriptional regulation integrates molecular events, such as the assembly of transcriptional machinery and RNA polymerase II (Pol II) binding, with physical processes like chromatin folding and steric hindrances, across a broad range of spatiotemporal dimensions. Yet, the precise mechanisms bridging these scales remain elusive. To initiate transcription, transcription factors (TFs) and other transcription effectors must navigate large distances in the crowded nuclear environment to encounter CREs. Spatial organizational strategies, including condensates and transcriptional hubs, have emerged as conserved mechanisms to concentrate transcription effectors, facilitating the assembly of active gene loci and bridging spatial scales [1,6–10].

Once transcription effectors reach their targets, the challenge extends beyond transcriptional initiation to coordinating molecular dynamics with the timescales of transcriptional bursts. TF binding events typically last only seconds, whereas transcriptional bursts can persist for several minutes [1,11,12]. Insights into this disparity have emerged from studies in promoters with multiple TF binding sites, where cooperative binding through continuous TF exchange maintains high promoter occupancy, effectively bridging the shorter timescales of TF binding with the longer duration of transcriptional bursts [13,14] (Figure 2A).

Beyond local molecular interactions, E-P communication plays a central role in gene regulation [1,2] (Figure 1A). Although various models, including both direct contact and action-at-a-distance mechanisms, have been proposed to explain E-P communication [1], the dynamic and transient nature of chromatin configurations complicates this understanding [15]. Recent studies tracking chromatin dynamics have begun to clarify this picture. Fully extruded loops, which group gene regulatory units, are relatively rare and short-lived [16,17]. For example, tracking the boundaries of a silent TAD in mouse embryonic stem cells (mESCs), a fully extruded loop was observed only about 3% of the time, persisting for 10 to 30 min, whereas a partially extruded configurations were more common, occurring 92% of the time [16,17]. The constraints imposed by cohesin and CTCF on extruded regions may help stabilize interactions and facilitate productive E-P communication [17]. However, whether productive E-P communication extends throughout the entire transcription process or occurs more rapidly remains an open question [18]. Notably, the relationships between E-P contact probabilities and transcription follows a non-linear, sigmoidal pattern, suggesting the involvement of intermediate regulatory steps bridging short-lived E-P contacts and transcriptional bursts [18] (Figure 2B). These regulatory steps, while abstract may represent a combination of stochastic regulatory processes, including TF recruitment, Mediator assembly, pre-initiation complex formation, and Pol II pausing and release, all contributing to transcriptional activity [18].

Live-cell imaging further indicates that sustained proximity, rather than direct physical contact, between E-P pairs is crucial for transcriptional activity. E-P pairs are typically 150–300 nm apart [19–24], and transcriptional output drops significantly when this proximity is disrupted [19,20]. However, the molecular signals transmitted during E-P interactions remain elusive, particularly given the small size of transcription complexes (10–20 nm) [1,25].

Emerging evidence suggests that transcriptional hubs, condensates, and clusters facilitate E-P communication by concentrating transcription effectors at gene loci, which can influence transcriptional bursting dynamics [1,6,8,22,26,27] (Figure 2C). These bursts can be regulated at the level of the OFF periods (burst frequency) and/or ON periods (burst size), thus modulating transcriptional activity [28]. For instance, embedding promoters within enhancer-associated clusters of Pol II general transcription factors (GTFs) has been shown to increase bursting frequency. The overlap between E-P interactions and GTFs activities suggest a mechanism for how nanoscale clusters extend productive E-P communication beyond molecular-scale interactions [22]. Additionally, condensates have been implicated in super-enhancer-driven gene bursting, where their assembly may help facilitate E-P communication [26]. E-P communication has been proposed to follow dynamic contact models, which describe productive E-P interactions as occurring in two phases: a brief E-P encounter phase followed by dissociation (e.g., "hit-and-run", "kiss-and-kick"), modulating transcriptional activity [1,10,26,29]. In mESCs, transcriptional burst size and frequency increase when Pol II condensate is spatially proximal (<1 μm) to the enhancer and gene locus ("three-way kissing"), with transcriptional activity decreasing as the condensate moves away. Further, depletion of cohesin does not affect basal transcription but limits burst size amplification when condensates are proximal [26], highlighting a complex regulatory network involving transcriptional hubs, condensates, and chromatin regulators. Interestingly, these mechanisms may follow general principles, as proposed by real-time measurements of endogenous transcriptional bursting in *Drosophila*: changes in gene activity levels are mainly linked to modulation of burst frequency at low-transcribing regime and burst size at mid-to-high transcribing regime [28].

Transcriptional hubs and condensates do not remain static [12]. In *Drosophila*, hubs mature progressively through compositional changes into Pol II-enriched clusters that drive transcriptional activation. Each compositional transition is both selective and tightly regulated, determining the number of active hubs [30,31]. Notably, these hubs are transient and dispersive, with transcription itself acting as a negative feedback mechanism that limits burst duration by dispersing hub components [30]. For instance, in zebrafish, transcriptional elongation can release E-P contacts, further emphasizing the dynamic nature of these structures [29]. It has also been proposed that hubs can retain a compositional footprint, which may serve as a 'molecular memory' to facilitate rapid reactivation [31].

Transcriptional activity also influences larger chromatin structures at the gene locus (Figure 2D). Locus decompaction is commonly associated with transcriptional activity, though its relationship with transcription varies across species. In yeast, highly transcribed genes tend to be less compact [32]. In contrast, mammalian systems show positive correlations between gene activity and locus compaction [33]. Fixed-imaging studies in *Drosophila* embryos suggest that inactive loci are more compact, while active loci show spatial decompaction with enhancers physically closest to the promoter [21]. These findings align with the observation that transcription-associated topologies, such as promoter-promoter and E-P contacts, are more prominent within highly active loci [21,33]. Chromatin decompaction likely facilitates genome accessibility, allowing regulatory elements to reorganize into proximity. This process could stabilize a more confined locus configuration, potentially reinforced by transcription (Figure 2D). Accordingly, loci with paired E-P often exhibit smaller E-P separations when transcriptionally active compared to their silent state, where dissociation occurs more rapidly [20]. Some models propose that transcriptional hubs and condensates can promote gene bursting even when E-P distances increase, potentially reflecting phase separation events or active spacing mechanisms during transcription [7,10,24].

Connecting the spatiotemporal parameters of molecular events, transcription, and E-P dynamics is crucial for uncovering general principles of transcriptional regulation. Key questions include: How frequently do E-P pairs form productive interactions? How long do these interactions last? How do they relate to transcription bursts and locus conformation? Clarifying these mechanisms will be essential for understanding how genetic information is transmitted, integrated, and regulated at the gene locus level.

**Chromatin Dynamics Across Scales: Polymer Models as Bridging Tools**

At the gene locus scale, transcriptional regulation relies on the coordinated integration of molecular signals and E-P interactions. However, a complete understanding requires examining larger organizational scales, particularly how genes and regulatory elements are positioned within the genome. This section focuses on chromatin dynamics and how spatial organization impacts the search process for regulatory interactions, ultimately shaping transcriptional control.

Certain enhancers, such as the ZRS enhancer located approximately 850 kb from the Shh promoter, can regulate gene expression over vast genomic distances. However, enhancer relocation studies have demonstrated significant distance-dependent effects on transcriptional activation. For instance, as the genomic separation between enhancer and promoter increases, interactions may occur less frequently or with limited functionality [18,19,34–37]. The probability of E-P contact is proposed to follow a non-linear, sigmoidal relationship with transcriptional output [5,18], where genomic distance influences the frequency of interactions (interburst duration) but has minimal impact on burst size or duration [18]. This non-linearity helps reconcile the statistical nature of TADs with the moderate transcriptional control exerted by contact probabilities, as recently reviewed [38].

Interestingly, enhancer activity over long distances may be modulated by specific regulatory sequences distinct from those governing cell-type specificity. When such long-range elements are incorporated, short-range enhancers can extend their functional range [35]. Several studies suggest that enhancers possess an intrinsic "action radius" or "sphere of influence," governed by their regulatory strength, which can be expanded by specific regulatory DNA sequences [35] or by forming cooperative networks with other enhancers in multi-enhancer clusters [34]. These mechanisms may allow enhancers to bypass some spatial constraints and gain flexibly to control gene activity over varying distances.

Beyond sequence-based regulation, dynamic processes shaping chromatin architecture also influence long-range regulatory interactions. Mechanisms such as cohesin- and CTCF-mediated loop extrusion help organize chromatin into spatially favorable configurations, bringing distant regulatory elements into proximity [36,39]. Understanding chromatin's physical organization and dynamics is therefore crucial for elucidating how distal regulatory elements regulate gene expression.

A central challenge in chromatin biology is understanding how a polymer as long as the genome (centimeters to meters) folds into a nucleus only 5–20 micrometers in diameter. Chromatin achieves this through a complex interplay of physical forces and biochemical interactions, forming distinct functional domains while remaining dynamic. To address this complexity, polymer physics models and simulations have proven valuable by providing simplified but predictive frameworks to describe chromatin behavior ([40,41], Table 1).

Polymer models typically reduce molecular detail while preserving essential physical properties (coarse-grained). They can be divided into two main classes: data-driven models, which reconstruct 3D chromatin conformations based on experimental data with minimal assumptions, and mechanistic models, which incorporate physical principles and specific hypotheses about chromatin organization, such as loop extrusion and phase separation. Key polymer physics metrics derived from imaging and sequencing studies help characterize chromatin behavior quantitatively: (1) physical distance, $R(s)$: the average separation between two loci at genomic distance $s$; (2) contact probability, $P(s)$: the likelihood of physical interaction between loci separated by $s$; (3) mean squared displacement ($MSD(t)$): the average movement of chromatin segments over time ($t$), reflecting chromatin diffusion and mobility. Comparing these metrics with polymer models such as the ideal chain, self-avoiding walk, equilibrium globule, and fractal globule can reveal how chromatin is organized and moves within the nucleus (reviewed in [42–45]; Figure 3A).

In a human cell line, contact probability measurements showed an exponent of $α ≈ 1$, which is not found in equilibrium polymer states, but fits a fractal globule polymer—a knot-free, crumpled polymer state where chromosomes segregate into compact territories and limit interchromosomal contacts (Figure 1B) [46–48]. Subsequent studies have shown $α$ to be system-dependent, varying between different chromosomes and species [49].

Simultaneously, chromatin exhibits dynamic properties consistent with the Rouse polymer model, describing chromatin as a flexible, entropic chain driven by thermal fluctuations and entropy [15–17,19,50,51]. This apparent paradox—where chromatin exhibits both fractal compaction and Rouse-like dynamics—was highlighted in recent studies in *Drosophila* embryos. Live imaging of ectopic E-P pairs across genomic distances (50 kb to 3 Mb) revealed that physical distances and interaction probabilities scaled with genomic distance according to the fractal globule model, while *MSD* measurements followed the predictions of an ideal Rouse chain [19]. This anomalous combination of compaction and dynamics has profound implications for transcriptional regulation. For E-P pairs separated by ~3 Mb, the DNA polymer's relaxation time (the time for the polymer to reconfigure) was measured to be 100 times shorter than predicted by either model alone, suggesting active mechanisms might assist long-range E-P interactions in driving gene regulation.

Multiple live imaging studies have consistently shown that chromatin behaves according to the Rouse model ([15–17,19,50,51], Figure 3B). These studies are beginning to shed light on both the slow and fast structural changes of chromatin, depending on the spatial scale under investigation (Figure 3B). For example, a 4 Mb subtelomeric region in human cells exhibited subdiffusive motion ($α ~ 0.5$), being only moderately constrained by nuclear factors such as chromatin interfaces and nuclear landmarks when displaced magnetically [50]. Meanwhile, smaller regions (~8 kb) randomly inserted into mESCs showed similar subdiffusive behavior but with higher effective diffusion coefficient ($D$), suggesting faster local chromatin movement [17,50]. In *Drosophila* embryos, 5 kb regions separated by ~550 kb exhibited similar dynamics (Figure 3B) [19]. However, tracking the boundaries of a 550 kb TAD in mESCs revealed more restricted motion potentially due to loop extrusion, with $D$ values 20-fold lower [16]. Such variations indicates that mobility is strongly influenced by the genomic context in question [52], such as differences in the processes underlying CTCF boundary interactions and E-P search, but might also reflect effects of transcription on chromatin mobility or species-specific differences.

Within TADs, physical constraints imposed by loop extrusion and chromatin compaction can influence regulatory dynamics by stabilizing regulatory interactions like E-P pairing. For example, tracking 8 kb regions within a neutral TAD in mESCs revealed constrained subdiffusive motion ($α ~ 0.2$) (Figure 3B) [17]. However, theoretical models suggest that as genomic regions approach sub-kilobase scales, their mobility may either increase or decrease depending on local chromatin architecture and biochemical interactions [15,17] (Figure 3C). While chromatin can display both slow and fast dynamic changes, the behavior of chromatin at intermediate scales, such as TADs, remains poorly understood (Figure 3D). These dynamics are critical to transcriptional regulation, as they likely influence how enhancers and promoters encounter each other and initiate transcription.

Integrating classical polymer physics models with live-cell imaging and genome-wide assays has proven invaluable for describing chromatin behavior (Table 1). However, capturing both local and global chromatin organization, while integrating E-P dynamics and transcriptional activity, remains a significant challenge. Moving forward, combining high-resolution imaging, advanced polymer simulations, and quantitative models will be crucial for dissecting the multiscale regulation of chromatin during gene expression (Figure 3D).

## Coordinating Genome Organization with Developmental Timing

Development is an intricately choreographed multicellular process requiring cells to integrate diverse signals and coordinate genomic activities with remarkable precision to guide tissue patterning and morphogenesis. While previous sections have focused on gene regulation at the individual cell level, development requires scaling up to the tissue level, where genomic processes must be synchronized and regulated across groups of cells. Tracking chromatin dynamics across different cell types is crucial for understanding how chromatin organization influences cell fate decisions and how cells diversify during development.

Single-cell atlases of *Drosophila* early embryogenesis, integrating chromatin accessibility (ATAC-Seq) and gene expression (RNA-Seq), have revealed that chromatin accessibility often precedes gene expression. This suggests a "priming" mechanism where regulatory regions become accessible before transcription initiates [53]. Consistent with this, loops and CRE hubs were observed to form early in cells with distinct fates, preceding both TAD formation and transcriptional activation [54,55].

In mouse embryos, most post-gastrulation cells exhibit mild heterogeneity in chromatin conformation at the TAD and A/B compartment levels. However, terminally differentiated cells, such as primitive erythrocytes, display compact, highly organized chromatin folding and strongly demarcated compartments [56]. This suggests that large-scale chromatin architecture remains relatively stable during early development, maintaining cellular plasticity, while finer-scale chromatin features, such as E-P contacts, may play a more decisive role in cell fate decisions and lineage specification [56–60].

Chromatin in pluripotent cells appears primed for flexibility. Many enhancer regions in these cells are pre-accessible, suggesting the genome is strategically organized to support a range of potential developmental pathways [59], This flexible regulatory landscape allows for rapid responses to differentiation cues while preserving developmental plasticity.

As differentiation progresses, chromatin structure becomes more specialized. Higher-order chromatin in undifferentiated mESCs is less compact, more dynamic, and more homogenous than in differentiated neuronal cells, which exhibit greater structural and dynamic heterogeneity [61]. A recent in vivo study quantified this plasticity by measuring the viscoelastic properties of chromatin, revealing that undifferentiated chromatin behaves as a Maxwell fluid, while differentiated chromatin displays both fluid-like and solid-like phases [61]. These findings align with earlier observations of a biphasic chromatin organization, where euchromatin and heterochromatin contribute to transcriptional activity and compaction, respectively [62].

Furthermore, euchromatin activity in differentiated states has been shown to enhance this biphasic continuum by promoting chromatin segregation and heterochromatin compaction, potentially stabilizing nuclear architecture [63]. While these associations between chromatin architecture and differentiation states are increasingly clear, the extent to which chromatin structure drives or responds to differentiation remains an open question.

Two pivotal processes exemplify how development bridges molecular and organismal scales: zygotic gene activation (ZGA) and gene patterning. Both require the coordination of chromatin structure with gene regulatory networks across spatial and temporal scales.

ZGA marks a critical transition in early development when external maternal signals trigger genome-wide transcription and chromatin reorganization across the embryo. This process synchronizes chromatin dynamics and gene expression across all embryonic cells. The earliest chromatin compartmentalization signals during ZGA typically emerge at the TAD level, a pattern conserved across species. However, the timing of features such as A/B compartment formation and loop extrusion can vary significantly across species [64]. For example, A/B compartments appear later in humans than in mice and *Drosophila*, while loop formation timing also diverges between species.

Despite such conservation of regulatory principles, cross-species comparisons reveal substantial variation, likely due to differences in genome size, developmental timing, and experimental resolution. These differences highlight how chromatin structure and transcriptional activation remain tightly linked but context-dependent during ZGA.

Gene patterning further illustrates how development bridges scales. In *Drosophila*, the Bicoid morphogen gradient initiates a cascade of transcriptional responses across the embryo, starting from nuclear-level transcriptional activation. The localized transcriptional readouts of Bicoid are progressively integrated, producing a step-like transcriptional gradient of hunchback, a downstream transcription factor that regulates subsequent developmental processes [65,66].

This gradual integration of transcriptional activity ensures coordinated patterns of gene expression, directing body segmentation and tissue formation across developmental timeframes. As development progresses, these patterns are further refined by cross-regulation between target genes and feedback mechanisms, reinforcing the stability of transcriptional programs.

Post-transcriptional regulation is a critical layer of gene expression control across scales. Notably, RNA metabolism—including transcript diffusion, localization, and fate—can contribute significantly to patterning and morphogenesis. In many organisms, transcription is halted prior to fertilization, making early developmental decisions rely on the post-transcriptional control of maternal mRNAs preloaded into the cytoplasm, before the onset of zygotic transcription. Additionally, RNA diffusion and localization within the cytoplasm can modulate transcript distribution and local translation, facilitating cross-regulation and feedback loops. For example, localized mRNA translation is involved in processes such as synapse formation and asymmetric cell fate decisions [67,68]. These mechanisms help spatially restricted gene products influence tissue-wide patterning, further linking molecular events to organism-scale outcomes.

Taken together, these examples illustrate how development achieves precise coordination of gene regulatory networks across multiple spatial and temporal scales. From the priming of chromatin accessibility in pluripotent cells to the genome-wide transcriptional reprogramming during ZGA and the orchestration of gene expression patterns during tissue patterning, chromatin architecture emerges as a dynamic regulatory scaffold.

These integrated mechanisms, which span from nuclear dynamics to tissue-level coordination, underscore how complex multicellular organisms emerge from the synchronized action of individual cells. Moving forward, investigating how chromatin structure and transcriptional activity reinforce each other across scales will be essential for a deeper understanding of developmental biology.

## Discussion and future perspectives

### Challenges for modeling scale-dependent chromatin dynamics

Chromatin displays a broad spectrum of dynamic behaviors, from rapid local reorganizations occurring within seconds to minutes [13–15,19,20] to slower, large-scale structural changes unfolding over hours to days [3,54,56]. Unraveling the mechanisms bridging these fast and slow timescales remains essential for linking chromatin structure with transcriptional regulation and gene expression control.

The interplay between gene loci dynamics and chromatin reorganization highlights the complexity of transcriptional regulation across spatial and temporal scales. Transcriptional activity can influence chromatin folding within TADs [15,20,21,33,69] and near TAD boundaries [70–75]. In turn, chromatin organization impacts transcription through mechanisms such as tethering elements and insulators [76,77]. Disruptions in these elements can alter transcriptional regulation, though effects often appear limited to local chromatin structures [77]. This emphasizes the challenge of reconciling higher-order chromatin structure with dynamic transcriptional activity, especially given observations of large-scale chromatin structure appearing decoupled from gene expression output [33,54,78–81].

Despite progress in polymer models of chromatin organization, no single theoretical framework accurately describes all aspects of chromatin behavior across scales, and polymer models that incorporate transcription are only starting to emerge [51]. Existing models often struggle to incorporate both the dynamic aspects of transcription and large-scale structural changes. Leveraging polymer physics to develop models that integrate global chromatin dynamics, local CRE activity, and transcriptional bursts remains a key area for exploration.

A significant barrier to generalizing chromatin behavior lies in variations across chromatin regions and experimental contexts. Dynamic behaviors can differ depending on region size, simulation scale, and local chromatin properties such as loop density, chromatin accessibility, and macromolecular crowding [15,17,82,83]. For example, recent loop extrusion polymer models showed how simulating different looping mechanisms, topological (where cohesin embraces DNA) and non-topological (cohesin acts as a cross-linker), impacts the observed contact probabilities and polymer behaviors [83,84]. In the non-topological model, loops form a comb-like polymer structure with a rigid central backbone and short, unentangled loops [83]. Conversely, the topological model preserves a linear polymer structure with more flexible chromatin loops [84].

The observed discrepancies between chromatin's structural and dynamic properties underscore the limitations of purely scale-free polymer models [85]. Incorporating the viscoelasticity of the nucleoplasm [85,86] and refining heteropolymer models to account for variation along the chromatin fiber, transcription activity, and regulatory interactions could enhance model accuracy. Mechanistic polymer models that integrate these properties could generate testable predictions, guiding experimental strategies to better capture chromatin's multi-scale behavior.

### Insights into phenotypic and species complexity

The complexity of chromatin architecture highlights that phenotypic diversity and species complexity arise not solely from genomic expansion but from the diversification and refinement of gene regulatory strategies. Though gene expression principles appear highly conserved, species-specific differences in genome organization can significantly influence chromatin structure and transcriptional dynamics. For instance, substantial variation in chromosome size and number between species might influence chromatin compaction mechanisms. Differences in cell cycle duration, particularly during early developmental stages, can further affect or be the consequence of the temporal dynamics of chromatin organization and transcriptional regulation.

Metazoan genomes rely on a combination of loop extrusion and focal DNA-DNA contacts. Vertebrates tend to rely more on loop extrusion [3,76,77], potentially resulting in a more linear chromosomal organization with slower relaxation dynamics due to persistent long-range correlations [83,84]. In contrast, invertebrates like flies exhibit more focal contacts [76,77], which may lead to a comb-like chromatin organization at smaller scales, allowing for faster domain dynamics as predicted by loop extrusion simulations [83,84].

Throughout development, cells must synchronize rapid changes in genome architecture with transcriptional programs. Elements such as CTCF binding and cohesin-mediated extrusion likely act as temporal regulators, facilitating genome compartmentalization while preserving nuclear organization. Species-specific differences in these regulatory mechanisms could influence the timing of chromatin reorganization and developmental progression, adding a layer of complexity to cross-species comparisons of gene regulation.

*Concluding Remarks and Open Questions*

Despite substantial progress, several critical questions remain regarding the interplay between chromatin dynamics and gene regulation:

- Plasticity and Precision: How do chromatin dynamics maintain both regulatory flexibility and precise developmental coordination across scales?
- Temporal Control: What additional mechanisms contribute to the temporal regulation of chromatin and transcriptional programs?
- Species-Specific Variation: How do species-specific differences in genome organization impact regulatory complexity and developmental timing?
- Feedback Mechanisms: To what extent do transcriptional bursts actively reshape chromatin architecture?

Addressing these questions will require continued exploration of both experimental and theoretical approaches that capture the temporal and spatial dimensions of chromatin regulation during development. Advancing multi-scale models, while integrating insights from polymer physics, will be critical for uncovering generalizable principles of gene regulation across species.

**Declaration of Competing Interest**
The authors declare no conflict of interest.

**Acknowledgments**
We thank D. Brückner and B. Zoller for careful reading and comments on the manuscript.

This work was supported in part by the European Research Council grant DynaTrans (101118866); by U.S. NSF, through the Center for the Physics of Biological Function (PHY–1734030); by U.S. NIH Grants R01GM097275, U01DA047730, and U01DK12742

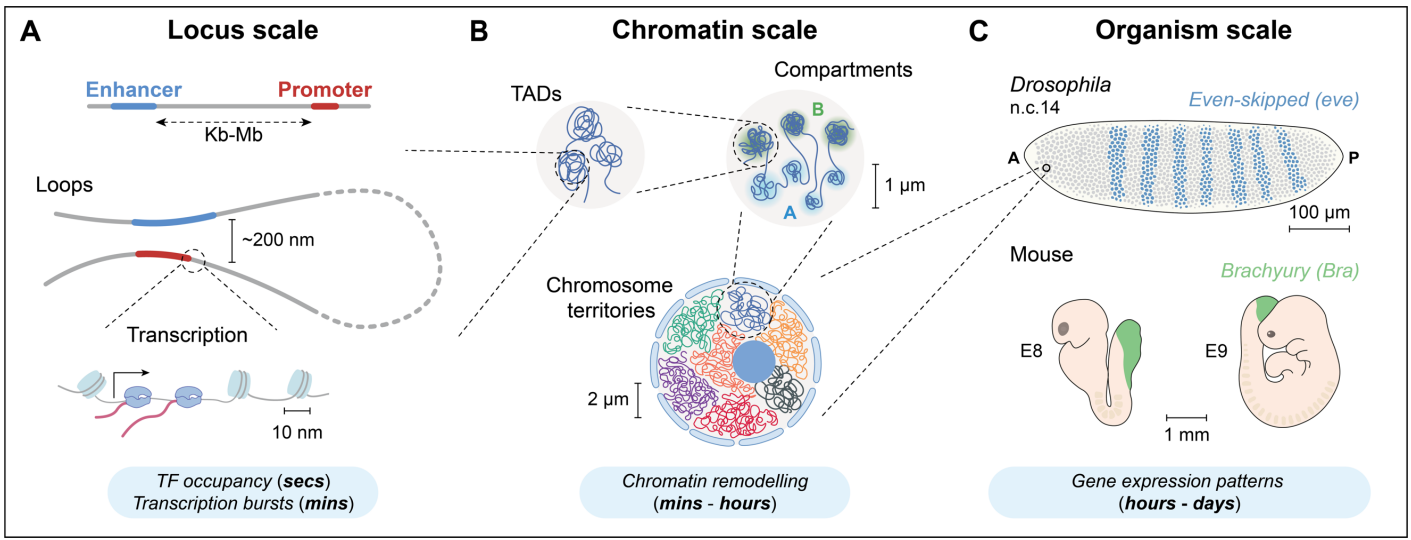

**Figure 1. Spatiotemporal scales of gene regulation during development.**
(**A**) **Enhancer-Promoter (E-P) interactions at the locus scale.** Enhancers (blue) and their target promoters (red) interact dynamically through loops that bring them into close proximity (~200 nm) but not direct contact. These E-P loops facilitate transcription initiation, typically occurring in short, stochastic bursts lasting minutes. Transcription factor (TF) binding, which occurs on even shorter timescales (seconds), contributes to transcriptional activation. Together, these rapid interactions establish the first layer of gene regulation, localized at individual loci. (**B**) **Higher-order chromatin structures at the chromatin scale.** Chromatin is organized into topologically associating domains (TADs), which are megabase-sized regions of increased self-interaction. TADs are further segregated into A and B compartments, reflecting differences in genomic activity and accessibility. Chromosomes occupy distinct, non-interlaced nuclear territories. Chromatin reorganization at this scale occurs over minutes to hours, enabling dynamic adjustments in gene regulation. (**C**) **Gene expression patterns at the organism scale.** The coordinated activity of E-P interactions at the locus scale and the dynamic restructuring of chromatin at the chromatin scale together generate spatially and temporally regulated gene expression patterns across tissues and developmental stages. This coordination unfolds over extended timescales, from hours to days, synchronizing gene expression programs across multiple loci and cell types. Shown are examples of the *even-skipped* (*eve*) gene in the *Drosophila* embryo at nuclear cycle (n.c.) 14 and the *Brachyury* (*Bra*) gene in the mouse embryo at embryonic (E) days E8 and E9, where gene activity patterns guide tissue formation and body axis specification.

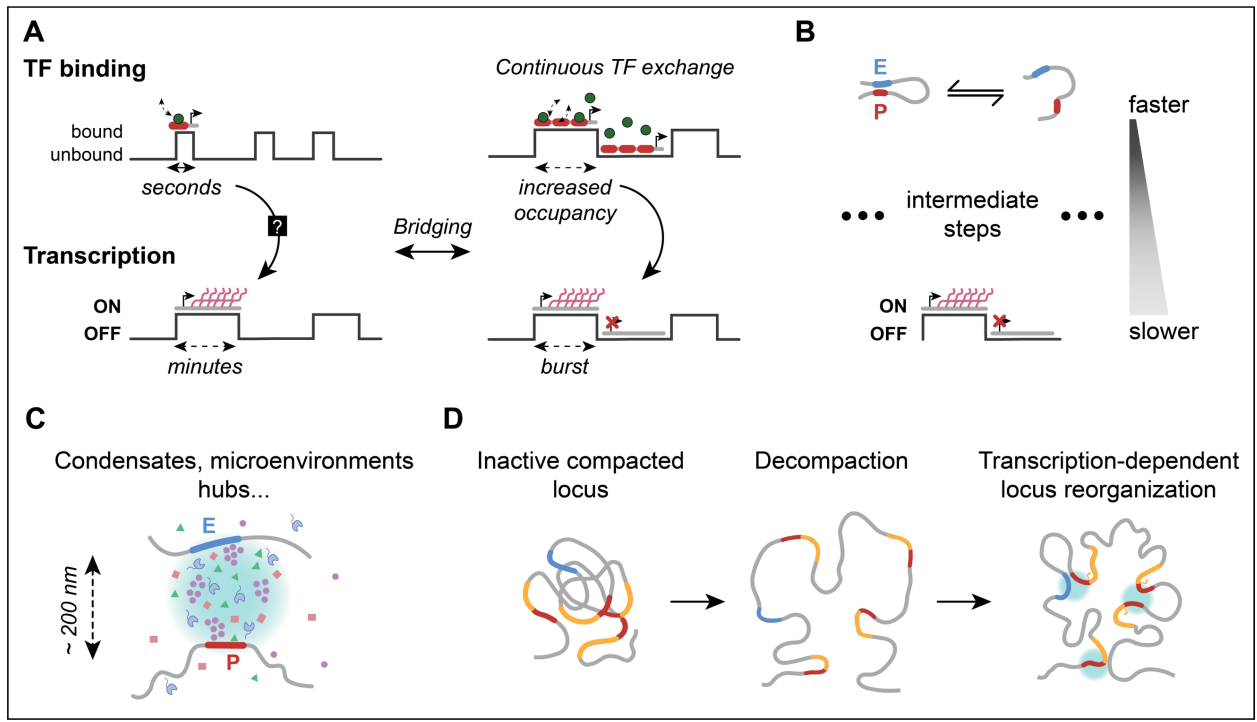

**Figure 2. Mechanisms bridging temporal and spatial scales in transcriptional regulation.**
(**A**) **Cooperative transcription factor (TF) binding and transcriptional bursts.** TFs bind and unbind promoters on short timescales (seconds). However, transcriptional bursts typically persist for longer periods (minutes). Continuous TF exchange, particularly at promoters with multiple binding sites, maintains high promoter occupancy, bridging short binding events with transcriptional bursts timescales. Adapted from Pomp et al. [13]. (**B**) **Enhancer-promoter (E-P) interactions and regulatory steps.** Zuin et al. [18] observed a sigmoidal (nonlinear) relationship between E-P contact probabilities and transcription levels, rather than a direct linear correlation. This finding suggests that physical E-P proximity alone does not directly translate to transcriptional output. Instead, a series of intermediate regulatory steps, such as TF recruitment, Mediator complex assembly, pre-initiation complex formation, and RNA polymerase pausing/release, collectively modulate transcription activity. These steps act intervene in the integration of rapid E-P interactions with slower transcriptional bursts. (**C**) **Condensates and transcriptional hubs.** Transcriptional condensates and microenvironments (~200 nm) concentrate TFs, coactivators, and RNA Pol II, potentially mediating E-P proximity and facilitating the assembly of active transcriptional machinery. Such hubs may influence transcriptional bursting dynamics. (**D**) **Transcription-dependent chromatin reorganization.** Hypothetical model illustrating transcription-dependent chromatin decompaction and reorganization. In the inactive state, the locus is highly compacted, limiting regulatory interactions. Chromatin decompaction enhances genome accessibility and allows regulatory elements to come into closer proximity. Transcription may subsequently stabilize a more confined and reorganized chromatin configuration.

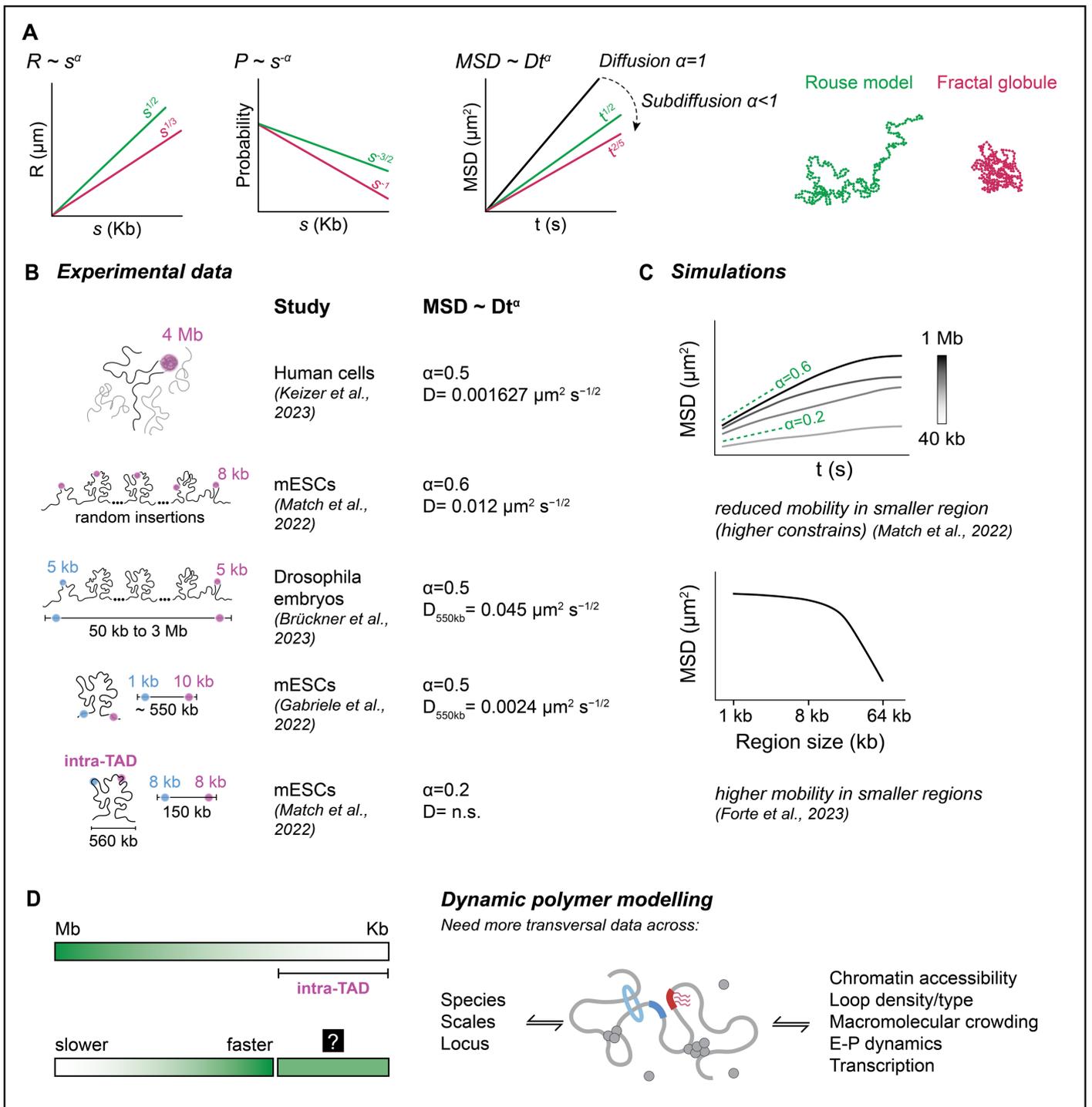

**Figure 3. Integrating chromatin dynamics across spatial scales.**
(**A**) **Scaling relationships predicted by polymer models**. The end-to-end distance (*R*), contact probability (*P*), and mean square displacement (*MSD*) are shown as functions of genomic distance (*s*) and time (*t*) for two widely used polymer physics models: the Rouse model (green) and the fractal globule model (red). The scaling coefficients for each model are indicated, highlighting the differences in chromatin behavior predicted by these theoretical frameworks. *D* in the *MSD* equation represents the diffusion quotient, which reflects the average rate of loci movement and the distance moved over time. **(B) Experimental measurements of chromatin mobility**. Summary of studies measuring chromatin dynamics using different cell types and genomic contexts. Diffusion exponents (*α*) and coefficients (*D*) vary across systems. On the left, the genomic segment sizes tracked in each study are indicated. Studies that simultaneously tracked two regions (e.g., enhancer-promoter pairs or TAD boundaries) are represented with blue and violet dots. Variations in *α* and *D* reveal scale-dependent chromatin dynamics, but differences in experimental methodologies and genomic contexts caution against overgeneralization. (**C**) **Simulations of chromatin mobility across scales.** The top panel shows simulated *MSD* curves for different genomic distances (from 40 Kb to 1 Mb) in a polymer with loop extrusion, where smaller distances exhibit higher *MSD* values and greater constraints. Adapted from Mach et al. [17]. The bottom panel shows *MSD* values for the center of mass of chromatin segments measured with different probe sizes, indicating higher mobility in smaller regions, as shown by Forte et al. [15]. **(D) Dynamic polymer modeling and** **chromatin behavior**. Experimental results suggest that larger chromatin regions exhibit slower dynamics, while smaller scales display more rapid movement. However, as scales approach sub-kilobase resolution, additional factors such as loop extrusion, enhancer-promoter interactions, transcription, and macromolecular crowding significantly influence chromatin mobility, making predictions more complex. This complexity, represented by the black box with a question mark, underscores the need for dynamic polymer models that incorporate multiple species, scales, and genomic contexts. A more comprehensive model would account for chromatin accessibility, loop density, folding mechanisms, molecular crowding, and transcriptional interference to capture chromatin behavior accurately.

References: Keizer et al., 2023 [50]; Mach et al., 2022 [17]; Brückner et al., 2023 [19]; Gabriele et al., 2022 [16]; Forte et al., 2023 [15]. n.s. – not shown.

**Table 1. Coarse-grained polymer simulation models.**

Summary of commonly used chromatin polymer models, emphasizing the diversity of approaches and the trade-offs between data-driven and mechanistic modeling strategies. All models described here can be implemented using a coarse-grained bead-and-spring polymer framework. Mechanistic models incorporate physical principles from polymer physics, such as the ideal chain or fractal globule models, while data-driven models focus on reconstructing chromatin conformations directly from experimental datasets. The choice of model influences both the spatial and temporal resolution, as well as the ability to capture dynamic chromatin behavior. Each approach offers different advantages, with mechanistic models providing insights into underlying physical principles, while data-driven models prioritize accuracy in reconstructing observed chromatin architectures.

| | *Model* | *Description* | *Strengths* | *Weaknesses* | *Data fitting?* | *Input* | *Output* | *Refs.* |
|---|---|---|---|---|---|---|---|---|
| **Mechanistic, agnostic, Inverse** | Maximum Entropy (MaxEnt) inference | Statistical approach used to infer 3D chromatin structure from experimental data | Makes minimal assumptions beyond the experimental constraints | Lack of mechanistic insights. Computationally intensive for large-scale modeling. Often does not capture the dynamic nature of chromatin | Yes | Contact probability maps | Simulated contact probability maps | [87,88] |
| **Mechanistic, Forward** | Block-copolymer | The chromatin fiber is modeled as a chain of beads, where beads of the same type have attractive interactions, while different types have repulsive or neutral interactions. These interactions drive the spontaneous formation of domains and compartments | Accounts for heterogeneity in chromatin properties. Can model large genomic regions or whole chromosomes | Oversimplification. Often does not account for dynamic changes in chromatin. Does not include protein-mediated interactions | No | Contact probability maps. Epigenetic modifications | Simulated contact probability maps | [89] |
| | Springs and Binders Switch (SBS) | Type of block-copolymer. The springs represent the chromatin fiber, modeled as a chain of beads, and the binders the protein complexes, that can bind to specific sites on the chromatin. The binders can form bridges between different regions, creating loops. The system can switch between different conformational states based on binder concentration and affinity. Equilibrium model | Predicts sharp transitions in chromatin states (phase transitions). Directly incorporates the role of binding proteins. Makes predictions at both local and global scales | Reduces complex protein-chromatin interactions to simple binding events. Does not account for loop extrusion dynamics. High sensibility to input-parameters | Yes or No | Contact probability maps | Simulated contact probability maps | [89–97] |
| | Loop extrusion (LE) | Extrusion factors are simulated as dynamic units that can bind chromatin and move along it. Non-equilibrium model. Variations have been developed to considering extrusion without an active, ATP-dependent motor, but driven by thermal diffusion of transcription-induced supercoiling | Explore features not captured by simple polymer models (e.g., CTCF orientation bias). Makes testable predictions about the dynamics of loop formation | Oversimplification. Primarily explains loop and TAD formation, not all aspects of chromatin organization. Computationally intensive for large genomic regions | No | CTCF/cohesin binding. Epigenetic modifications. DNA accessibility | Ensemble of 3D chromatin conformations. Simulated contact probability maps | [16,17,83,84,90,98,99] |
| | Highly predictive heteromorphic polymer (HiP-HoP) | Represents chromatin with different levels of compaction (heteromorphic), combined with diffusing protein bridges and loop extrusion | Accounts for heterogeneity in chromatin properties. Can account for loop extrusion dynamics. Can incorporate multiple mechanisms of chromatin organization | Computationally intensive. Requires integration of multiple types of experimental data | No | CTCF/cohesin binding. Epigenetic modifications. DNA accessibility | Simulated contact probabilities maps, FISH | [15,100] |